\documentclass[11pt]{article}
\usepackage[margin=1in]{geometry}
\usepackage[T1]{fontenc}
\usepackage{microtype}
\usepackage{graphicx}

\usepackage{mathtools,amssymb,amsthm}
\usepackage{enumitem}
\usepackage{stmaryrd} 
\usepackage{braket}

\usepackage[
  backend=biber,
  style=alphabetic,
  maxalphanames=5,  
  minalphanames=5,  
  maxbibnames=5,    
  minbibnames=5,    
  maxcitenames=5,
  mincitenames=5,
  doi=true,
  url=true,
  isbn=false
]{biblatex}

\usepackage{xcolor}
\usepackage{hyperref}
\hypersetup{
  colorlinks=true,
  linkcolor=blue,
  citecolor=green!50!black,
  urlcolor=cyan
}
\usepackage{aliascnt}
\usepackage[nameinlink,noabbrev]{cleveref}
\crefname{fact}{Fact}{Facts}
\Crefname{fact}{Fact}{Facts}
\crefname{remark}{Remark}{Remarks}
\Crefname{remark}{Remark}{Remarks}

\newtheorem{theorem}{Theorem}[section]
\newaliascnt{lemma}{theorem}
\newtheorem{lemma}[lemma]{Lemma}
\aliascntresetthe{lemma}
\newaliascnt{proposition}{theorem}

\aliascntresetthe{proposition}
\newaliascnt{corollary}{theorem}

\aliascntresetthe{corollary}
\newaliascnt{fact}{theorem}
\newtheorem{fact}[fact]{Fact}
\aliascntresetthe{fact}
\theoremstyle{definition}
\newaliascnt{definition}{theorem}

\aliascntresetthe{definition}
\newaliascnt{example}{theorem}

\aliascntresetthe{example}
\theoremstyle{remark}
\newaliascnt{remark}{theorem}
\newtheorem{remark}[remark]{Remark}
\aliascntresetthe{remark}
\newtheorem*{remark*}{Remark}

\DeclareMathOperator{\poly}{poly}
\DeclareMathOperator*{\expt}{\mathbb{E}}
\DeclareMathOperator*{\var}{\mathrm{Var}}
\newcommand{\ketbra}[2]{\lvert #1\rangle\!\langle #2\rvert}
\newcommand{\norm}[1]{\left\lVert #1 \right\rVert}
\newcommand{\abs}[1]{\lvert #1 \rvert}
\newcommand{\lrb}[1]{\llbracket #1 \rrbracket}
\newcommand{\clock}{{\rm C}}

\title{Vanilla Adiabatic Quantum Optimization Can Have Subexponential Advantage}
\author{Yufan Zheng\thanks{Visa Research. \url{yfzheng@visa.com}}}
\date{}

\begin{document}

  \maketitle

  \begin{abstract}
    We construct a family of objective functions $f \colon \{0,1\}^n \to \mathbb{R}$ for which vanilla adiabatic quantum optimization along $H(s) = -(1-s)\sum_{i=1}^n X_i + s\sum_x f(x)\ketbra{x}{x}$ finds the global minimizer in $\poly(n)$ time, whereas any classical algorithm requires $\exp(n^{\Omega(1)})$ time to do so in the black-box model.
    Our result is achieved by replacing the linearization technique in an earlier work of Leng et al. (arXiv:2504.14841), which establishes a similar subexponential quantum speedup using a transverse field Ising mixer rather than $-\sum_{i=1}^n X_i$.
  \end{abstract}

  \section{Introduction}\label{sec:introduction}

\emph{Adiabatic quantum optimization}~\cite{farhi2000quantum} solves a discrete optimization problem $\min_{x \in \{0,1\}^n} f(x)$ by initializing an $n$-qubit system in the ground state of a \emph{mixer} Hamiltonian $H_{\rm M}$ and evolving it under $H(s(\tau))$ for physical time $\tau$ and schedule function $s\colon [0,T] \to [0,1]$, where
\begin{equation} \label{eq:plain_path}
    H(s) = (1-s) H_{\rm M} + s \sum_{x \in \{0,1\}^n} f(x) \ketbra{x}{x}, \qquad s \in [0,1].
\end{equation}
If the spectral gap along this path does not close, the quantum adiabatic theorem~\cite{teufel2003, jansen2007bounds} guarantees that, for sufficiently large $T$, the system remains close to its instantaneous ground state and converges to the global minimizer of $f$.
In the original proposal of Farhi et al.~\cite{farhi2000quantum}, the mixer is equivalent to a simple transverse field $H_{\rm M} = -\sum_{i=1}^n X_i$.\footnote{This refers to the uniform-field choice studied there. Their general framework also allows unequal weights for different $X_i$.}
This approach has inspired extensive experimental work, with prominent examples in nuclear magnetic resonance systems~\cite{steffen2003experimental}, superconducting circuits~\cite{king2023quantum}, and Rydberg atom arrays~\cite{ebadi2022quantum}, to name a few.
Adiabatic quantum optimization has also motivated gate-based variational approaches, most notably the quantum approximate optimization algorithm (QAOA)~\cite{farhi2014quantum}.
Its simplicity and broad applicability make understanding its advantage over classical algorithms a fundamental question~\cite{Albash2018}.

Adiabatic quantum optimization was soon known to provide a quadratic speedup against classical algorithms for unstructured search~\cite{roland2002quantum}.
However, whether a superquadratic speedup is possible remained open for many years, even in the black-box model~\cite{vanDam2001,Aharonov2007,Albash2018}.
Stronger speedups were indeed known in computational models related to adiabatic quantum optimization:
Allowing a non-diagonal final Hamiltonian generalizes it to \emph{adiabatic quantum computation}, which is $\mathsf{BQP}$-complete when both endpoint Hamiltonians are local~\cite{Aharonov2007}. It therefore admits superpolynomial speedups unless $\mathsf{BQP} = \mathsf{BPP}$, in which case every quantum algorithm can be simulated classically with polynomial overhead.
Subexponential speedups over classical algorithms have also been established in the black-box model using diabatic transitions~\cite{somma2012quantum} or a detour through an intermediate Hamiltonian~\cite{gilyen2021subexponential}.
These results, however, did not establish superquadratic speedups for adiabatic quantum optimization along \eqref{eq:plain_path}.
It therefore remained possible that the algorithm could be simulated classically with only quadratic overhead.
Indeed, both theory and experiment show that \emph{classical} quantum Monte Carlo methods can reproduce tunneling effects in adiabatic quantum optimization with only polynomial overhead in certain settings~\cite{crosson2016simulated, denchev2016computational}.

To resolve this uncertainty, Leng et al.~\cite{leng2025subexponentialquantumspeedupoptimization} recently established a subexponential speedup for adiabatic quantum optimization in the black-box model.
They constructed a family of objective functions $f$ that adiabatic evolution along \eqref{eq:plain_path} minimizes in $\poly(n)$ time, while any classical algorithm requires $\exp(n^{\Omega(1)})$ evaluations of $f$.
The objective $f$ is obtained by compiling an earlier construction demonstrating a subexponential quantum speedup for stoquastic adiabatic computation by Gilyén, Hastings, and Vazirani~\cite{gilyen2021subexponential} through a sequence of perturbative reductions.
The Gilyén--Hastings--Vazirani construction in turn builds on Hastings' graph decoration technique~\cite{hastings2021power} and the welded-tree idea of Childs et al.~\cite{childs2003exponential}.

However, the subexponential quantum speedup in \cite{leng2025subexponentialquantumspeedupoptimization} relies on a mixer $H_{\rm M}$ that is a transverse field Ising Hamiltonian, so it does not cover textbook adiabatic quantum optimization with the plain transverse field mixer $H_{\rm M} = -\sum_{i=1}^n X_i$ used in Farhi et al.'s original proposal~\cite{farhi2000quantum} and most experimental implementations.

\paragraph{Our result.}
We close this gap: a subexponential speedup exists for the textbook adiabatic path.

\begin{theorem}[Informal version of \Cref{thm:main}]
\label{thm:intro}
    There exists a family of objective functions $f \colon \{0,1\}^n \to [-\poly(n),\poly(n)]$, accessed by an oracle, such that the following holds:
    \begin{itemize}
        \item Adiabatic quantum optimization along the path
        \begin{equation} \label{eq:intro_H_sigma}
            H(s) = -(1-s)\sum_{i=1}^{n} X_i + s \sum_{x \in \{0,1\}^n} f(x) \ketbra{x}{x}, \qquad s \in [0,1],
        \end{equation}
        finds the global minimizer $x^*$ in $\poly(n)$ time and queries to $f(\cdot)$.
        \item Any classical algorithm requires $\exp(n^{\Omega(1)})$ queries to $f(\cdot)$ to find an $x$ such that $f(x) - f(x^*) \leq 1/\poly(n)$.
    \end{itemize}
\end{theorem}

  \section{Technical Overview} \label{sec:idea}

Write $X_{[n]} \coloneqq \sum_{i=1}^n X_i$ for notational simplicity, where $X_i$ is the Pauli-$X$ operator acting on qubit $i$.

\paragraph{Starting point: a curved adiabatic path.}
Our starting point is the adiabatic path of \cite[Lemma 4.4]{leng2025subexponentialquantumspeedupoptimization}:
\begin{equation} \label{eq:intro_K}
    K(t) = -X_{[n]} + D(t), \qquad t \in [0,1],
\end{equation}
where $D(t)$ is diagonal, $D(0)=0$ and $D(1) = -3n\ketbra{u}{u}$ for a hidden string $u$ that is classically hard to find given oracle access to $D(t)$, and the path has a spectral gap $\Omega(1/\poly(n))$ throughout. 
By the quantum adiabatic theorem, evolving along $K(t)$ efficiently prepares the ground state of $-X_{[n]}+D(1)$. Measuring this state reveals $u$ with constant success probability, demonstrating a quantum speedup.
Comparing \eqref{eq:intro_K} with the target path \eqref{eq:intro_H_sigma} of \Cref{thm:intro}, we find two mismatches:
\begin{enumerate}[label=\arabic*.]
    \item The mixer in \eqref{eq:intro_K} is $-X_{[n]}$ but not $-(1-t)X_{[n]}$, so it never switches off: $K(1) = -X_{[n]} + D(1)$. In contrast, the target path must end at a diagonal Hamiltonian $H(1)$.
    \item $D(t)$ is only guaranteed to be Lipschitz in $t$, whereas the target path must be \emph{linear} in $s$.
\end{enumerate}

\paragraph{Linearization technique in \cite{leng2025subexponentialquantumspeedupoptimization}.}
We briefly revisit the construction of \cite{leng2025subexponentialquantumspeedupoptimization} that addresses the second mismatch above.
Write $\ket{c_j}\coloneqq\ket{1^j0^{k-j}}$.
The authors construct a clock Hamiltonian $H_\clock(s)$ that is \emph{linear} in $s$, such that increasing $s$ from $0$ to $1$ moves its ground state approximately through the sequence
\begin{equation}
    \ket{c_0}\longrightarrow
    \frac{\ket{c_0}+\ket{c_1}}{\sqrt2}\longrightarrow
    \ket{c_1}\longrightarrow
    \frac{\ket{c_1}+\ket{c_2}}{\sqrt2}\longrightarrow
    \ket{c_2}\longrightarrow\cdots\longrightarrow\ket{c_k}
\end{equation}
that transitions smoothly between successive computational states.
Now, consider a system consisting of a $k$-qubit clock register and an $n$-qubit problem register, with $k=\poly(n)$.
Construct
\begin{equation} \label{eq:intro_H_s}
    H(s)\coloneqq\Gamma H_\clock(s)\otimes I-I\otimes X_{[n]}
    +\sum_{j=0}^{k}\ketbra{c_j}{c_j}\otimes D(j/k).
\end{equation}
For sufficiently large $\Gamma$, the first term $\Gamma H_\clock(s)\otimes I$ dominates, so the low-energy states of $H(s)$ keep the clock close to the ground state of $H_\clock(s)$, approximately some $\alpha\ket{c_j}+\beta\ket{c_{j+1}}$ at any $s$. 
Therefore, for the last term in \eqref{eq:intro_H_s}, the problem register effectively sees $\abs{\alpha}^2D(j/k)+\abs{\beta}^2D((j+1)/k)$ by pinning the clock register to $\alpha\ket{c_j}+\beta\ket{c_{j+1}}$.
Consequently, while being \emph{linear} in $s$, $H(s)$ effectively ``simulates'' the path $-X_{[n]} + D(t) = K(t)$ on the problem register, preserving its spectral gap (for quantum easiness) and classical hardness.
The authors then apply a ``path tilt and extension'' technique to $H(s)$ to obtain a linear path that establishes the quantum--classical separation of \Cref{thm:intro} with $-X_{[N]}$ replaced by a transverse field Ising Hamiltonian.


\paragraph{Our approach: a tensor-power clock.}
Our idea is using a different clock mechanism, again with $N\coloneqq k+n$ qubits split into a $k$-qubit clock register and an $n$-qubit problem register.
Following the idea of the last paragraph, we want to control the clock register with a Hamiltonian $H_\clock(s)$ and use it to build a path similar to \eqref{eq:intro_H_s}. 
Our target form \eqref{eq:intro_H_sigma} of $H(s)$ basically restricts possible choices to
\begin{equation}
    H_\clock(s)=-(1-s)X_{[k]}+sW
\end{equation}
for some diagonal $W$.
We then choose $W \coloneqq \sum_{j=1}^k\ketbra{1}{1}_j$, where each projector acts on clock qubit $j$, i.e., $W$ counts the ones in a clock string.
Now, $H_\clock(s)$ is a $k$-fold Kronecker sum of a single-qubit Hamiltonian $A(s)=-(1-s)X+s\ketbra{1}{1}$.
The ground state $\ket{\psi(s)}$ of $A(s)$ rotates smoothly from $\ket{+}$ to $\ket{0}$, with magnetization
\begin{equation}
    m(s)\coloneqq\braket{\psi(s)|Z|\psi(s)}
    =\frac{s}{\sqrt{s^2+4(1-s)^2}},
\end{equation}
which increases continuously and strictly from $0$ to $1$.
It is then straightforward to derive that the clock has ground state $\ket{\psi(s)}^{\otimes k}$ and gap $\sqrt{s^2+4(1-s)^2}= \Omega(1)$.

Measuring this product state in the computational basis gives a string $c\in\{0,1\}^k$ of independent bits.
As the clock ground state moves from $\ket{+}^{\otimes k}$ to $\ket{0}^{\otimes k}$, we use the magnetization per qubit $t(c)\coloneqq1-2\abs{c}/k\in[-1,1]$ as a clock reading since its mean $m(s)$ advances from $0$ to $1$ as $s$ increases from $0$ to $1$:
\begin{equation}
    \expt_c[t(c)]=m(s), \qquad
    \var_c[t(c)]\leq\frac1k.
\end{equation}
In an idealized world, we want the clock to give a definite reading $t(c) = m(s)$ at each $s$, which is clearly not the case.
Nevertheless, increasing $k$ lets our clock approximate the ideal behavior as closely as needed, since its fluctuations shrink as $k$ grows.
For intuition, we can therefore pretend that the clock has reading $m(s)$ and is advancing from $0$ to $1$ as $s$ increases.

Now we construct $H(s)$ that has the target form \eqref{eq:intro_H_sigma}: similar to \eqref{eq:intro_H_s}, let
\begin{equation}
    \begin{aligned} \label{eq:constructed_H}
        H(s)\coloneqq H_\clock(s)\otimes I-(1-s)I\otimes X_{[n]}+s\widehat{G},\quad
        \widehat{G}\coloneqq\sum_{c\in\{0,1\}^k}\ketbra{c}{c}\otimes G(t(c)),
    \end{aligned}
\end{equation}
where $G(t)$ for $t\in[-1,1]$ is a diagonal problem Hamiltonian to be determined.
Pinning the clock to $\ket{\psi(s)}^{\otimes k}$, the ground state of $H_\clock(s)$, effectively lets the problem register see the last two terms in \eqref{eq:constructed_H} as $-(1-s)X_{[n]}+s\expt_c[G(t(c))]$.
In an idealized world, this equals $-(1-s)X_{[n]}+sG(m(s))$.
Recall that our goal is to let $H(s)$ effectively simulate $K(t) = -X_{[n]} + D(t)$ in \eqref{eq:intro_K}.
Therefore we can choose a $G$ such that
\begin{equation}
    -(1-s)X_{[n]}+sG(m(s)) = -(1-s)X_{[n]} + (1-s)D(m(s)) = (1-s)K(m(s)),
\end{equation}
which leads to
\begin{equation}
    G(t)=\lambda(t)D(t), \qquad
    \lambda(t)\coloneqq\frac{\sqrt{1-t^2}}{2t}, \qquad t\in (0,1],
\end{equation}
because $\lambda(m(s))=\frac{1-s}{s}$.

\begin{figure}[!htbp]
    \centering
    \includegraphics[height=0.2\textheight,trim={0 9pt 0 0},clip]{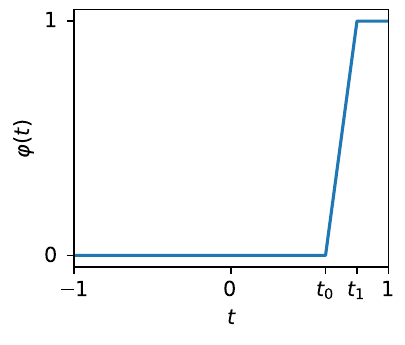}\hspace{0.1\textwidth}
    \includegraphics[height=0.2\textheight,trim={0 9pt 0 0},clip]{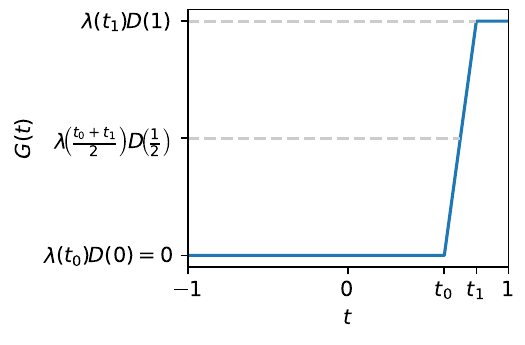}
    \caption{(Schematic) plots of $\varphi(t)$ and $G(t)$.}
    \label{fig:clock_reparametrization}
\end{figure}

\paragraph{Reparametrizing the clock reading.}
Three issues remain in the construction above:
\begin{enumerate}[label=\arabic*.]
    \item $\lambda(t) \to +\infty$ as $t\to0$, but $G(t)$ must be defined on $[-1,1]$.
    \item We have $\lambda(t(c))\to0$ as $t(c)\to1$ when $s \to 1$, which will close the spectral gap of $H(s)$.
    \item The $s\widehat{G}$ term in \eqref{eq:constructed_H} is not perturbatively small when $s$ is $\Omega(1)$-far from $1$, even if $\widehat{G}$ were bounded; therefore, the clock may not remain close to the ground state of $H_\clock(s)$.
\end{enumerate}
We resolve all three by fixing a small window $[t_0,t_1]\subset(0,1)$ sufficiently close to $1$, and defining $\varphi(t)$ to be $0$ for $t\leq t_0$, $1$ for $t\geq t_1$, and linear in between. Finally, set
\begin{equation}
    G(t)\coloneqq\begin{cases}
        0, & t \in [-1,t_0],\\
        \lambda(t)D(\varphi(t)), & t \in [t_0,t_1],\\
        \lambda(t_1)D(1), & t \in [t_1,1].
    \end{cases}
\end{equation}
It is easy to check that the new definition of $G$ still preserves the simulation mechanism described in the previous paragraph while resolving all three issues.
See \Cref{fig:clock_reparametrization} for (schematic) plots of $\varphi(t)$ and $G(t)$.

  \section{Proof}

This section proves \Cref{thm:main}. \Cref{sec:preliminaries} introduces notation and standard facts, \Cref{sec:construction} constructs $F_\sigma$, \Cref{sec:spectral_gap} bounds the spectral gap, and \Cref{sec:main_theorem} completes the proof.

\subsection{Preliminaries} \label{sec:preliminaries}

\paragraph{Sets and intervals.}
For a positive integer $n$, we use $[n]$ to denote the set $\{1,2,\ldots,n\}$.
For real numbers $\ell\le r$, we use $[\ell,r]$ to denote the set $\{t\colon\ell\le t\le r\}$.

\paragraph{Asymptotics.}
Throughout this paper, we use the standard asymptotic notations $O(\cdot)$, $\Theta(\cdot)$, and $\Omega(\cdot)$ to represent \emph{predetermined} expressions that do not depend on any other parameters.\footnote{For example, the statement ``if $f \geq O(g)$ then $a = \Theta(b)$'' means that there exist functions $g_0 = O(g)$ and $b_0 = \Theta(b)$ such that if $f \geq g_0$, then $a = b_0$.}
Additionally, we use $\poly(\cdot)$ to denote a function that is bounded below by $1$ and bounded above by some implicit polynomial in terms of the parameters within.
We emphasize that \emph{all} implicit functions represented in these asymptotic forms can be computed explicitly and efficiently.

\paragraph{Oracles.}
Our result will be proved in the \emph{oracle} model, a.k.a.\ the \emph{black-box} model.
The oracles in this paper will be classical functions that allow both classical and quantum access.

Formally, let $f\colon\{0,1\}^n\to\{0,1\}^m$ be a function, where we assume without loss of generality that the input and output of $f$ are binary strings.\footnote{We omit routine finite-precision encoding details for real-valued queries and answers, as polynomially many bits of precision suffice to preserve all asymptotic results in this paper.}
The oracle $O_f$ provides efficient query access to $f$:
\begin{itemize}
    \item in the classical case, $O_f$ directly evaluates $f$ on the classical query, i.e., $O_f(x)=f(x)$ for any $x\in\{0,1\}^n$;
    \item in the quantum case, $O_f$ handles queries in superposition and evaluates $f$ in superposition, i.e., $O_f\ket{x}_{\mathcal{A}}\ket{y}_{\mathcal{B}}=\ket{x}_{\mathcal{A}}\ket{y\oplus f(x)}_{\mathcal{B}}$ where $\mathcal{A}$ is the query register and $\mathcal{B}$ is the answer register; and $O_f\sum_{x,y}\alpha_{x,y}\ket{x}\ket{y}=\sum_{x,y}\alpha_{x,y}\ket{x}\ket{y\oplus f(x)}$ for any $\{\alpha_{x,y}\}_{x,y}$, where $\oplus$ denotes bitwise XOR.
\end{itemize}

We say $O$ is an oracle if $O=O_f$ for some function $f$.
We emphasize that the query oracle $O_f$ ignores the complexity of implementing $f$.

\paragraph{Diagonal Hamiltonians and Oracle Access.}
A Hamiltonian $H$ is called diagonal if it is diagonal in the computational basis.
In such a case, we sometimes regard $H$ as a function that maps computational basis states to their corresponding eigenvalues: $H\lrb{x} \coloneqq \braket{x|H|x}$.
Oracle access to a diagonal Hamiltonian $H$ means query access to $H$ as a function.

\paragraph{Norms, spectral gaps, Lipschitz constants, and adiabatic paths.}
We write $\norm{\cdot}$ for the operator norm of a matrix and the Euclidean norm of a vector.
If $A$ is Hermitian, let $\mu_0(A) \leq \mu_1(A) \leq \cdots$ denote its eigenvalues, counted with multiplicity, and $\delta_A \coloneqq \mu_1(A)-\mu_0(A)$ its spectral gap.
For a time-dependent Hamiltonian $A(t)$, $t \in I \subseteq \mathbb{R}$, define
\begin{equation*}
    M_A \coloneqq \sup_{t \in I} \norm{A(t)}, \qquad \delta_A \coloneqq \inf_{t \in I} \delta_{A(t)}.
\end{equation*}
We say $A(t)$ is $L$-Lipschitz in $t$ if $\norm{A(t)-A(t')} \leq L\abs{t-t'}$ for all $t,t' \in I$, and $L_A$ denotes the least such $L$.
An \emph{adiabatic path} is a time-dependent Hamiltonian intended for adiabatic evolution, typically with an inverse-polynomial spectral gap.

\paragraph{Pauli operators.}
We denote the standard single-qubit Pauli operators by $X$, $Y$, and $Z$. For multi-qubit systems, we use $X_i$, $Y_i$, and $Z_i$ to represent these operators acting specifically on qubit $i$.
For a positive integer $m$ we abbreviate $X_{[m]} \coloneqq \sum_{i \in [m]} X_i$.

\paragraph{Hamming weights.}
For a binary string $c$ we write $\abs{c}$ for its Hamming weight.

\medskip
We will use the following standard facts in subsequent subsections.

\begin{fact}[Weyl's inequality] \label{fact:weyl}
    Let $A,B$ be Hermitian operators on the same space.
    Then $\abs{\mu_j(A+B) - \mu_j(A)} \leq \norm{B}$ for every $j$; in particular $\delta_{A+B} \geq \delta_A - 2\norm{B}$.
\end{fact}

\begin{fact}[Kronecker sum] \label{fact:kron}
    Let $A_1,\ldots,A_m$ be Hermitian.
    Then the spectrum of the Kronecker sum $\sum_{i \in [m]} I^{\otimes(i-1)} \otimes A_i \otimes I^{\otimes(m-i)}$ is the Minkowski sum of the spectra of the $A_i$, with eigenvectors the corresponding tensor products of eigenvectors of the $A_i$.
\end{fact}

\subsection{Construction} \label{sec:construction}

In this subsection we construct the objective function $F_\sigma$ underlying \Cref{thm:main}.
Our starting point is a hard adiabatic path $K_\sigma$ from \cite{leng2025subexponentialquantumspeedupoptimization}:

\begin{theorem}[{\cite[Lemma 4.4]{leng2025subexponentialquantumspeedupoptimization}}] \label{thm:cited}
    For every $n$ in some infinite increasing sequence of positive integers with polynomially related consecutive terms, there exists a family $\{K_{\sigma}\}_{\sigma}$ of $n$-qubit time-dependent Hamiltonians, each of the form
    \begin{equation}
        K_{\sigma}(t) = - X_{[n]} + D_{\sigma}(t), \qquad t \in [0,1],
    \end{equation}
    where $D_{\sigma}(t)$ is diagonal, such that the following holds:
    \begin{enumerate}[label=(\alph*)]
        \item $D_{\sigma}(0) = 0$, $D_{\sigma}(1) = -3n \ketbra{u_{\sigma}}{u_{\sigma}}$ for some $u_{\sigma} \in \{0,1\}^n$, and $L_{D_{\sigma}} \leq \poly(n)$.
        \item There is a distribution over $\sigma$ such that, given oracle access to $D_{\sigma}(t)$ but no other information about $\sigma$, every classical algorithm making at most $\exp(n^{\Omega(1)})$ queries finds $u_{\sigma}$ with probability at most $\exp(-n^{\Omega(1)})$, over the draw of $\sigma$ and the algorithm's randomness.
        \item $\delta_{K_{\sigma}} \geq \frac{1}{\poly(n)}$ and $M_{K_{\sigma}} \leq \poly(n)$.
    \end{enumerate}
\end{theorem}

\begin{remark} \label{rem:normalization}
    \Cref{thm:cited} is a normalized restatement of \cite[Lemma 4.4]{leng2025subexponentialquantumspeedupoptimization}, whose parameter $n_{\mathsf{B}}$ we rewrite as $n$.
    The Hamiltonian $K_{\sigma}(t)$ here is obtained by dividing the Hamiltonian $H(t)$ there by $\Delta_{\mathsf{B}}$.
\end{remark}

Until \Cref{sec:main_theorem} we drop $\sigma$ from the notation, writing $K(t)$, $D(t)$, and $u$, as none of the reasoning in between depends on $\sigma$.
We also assume $n \geq 2$ without loss of generality.

\paragraph{Basic bounds.}
\Cref{thm:cited} gives
\begin{equation} \label{eq:cited_bounds}
    \delta_K \geq \frac{1}{\poly(n)}, \qquad M_K \leq \poly(n), \qquad L_D \leq \poly(n) .
\end{equation}
Since $D(t) = K(t) + X_{[n]}$ and $D(1) = -3n\ketbra{u}{u}$,
\begin{equation} \label{eq:MD_bounds}
    3n = \norm{D(1)} \leq M_D \leq M_K + \norm{X_{[n]}} = M_K + n \leq \poly(n) .
\end{equation}
All of these bounds hold uniformly in $\sigma$.
We may therefore fix a $\sigma$-independent quantity
\begin{equation} \label{eq:kappa_def}
    \kappa = \poly(n) \qquad\text{such that}\qquad \delta_{K_\sigma} \geq \frac{1}{\kappa}, \quad M_{D_\sigma} \leq \kappa, \quad L_{D_\sigma} \leq \kappa \quad \text{for every } \sigma.
\end{equation}

\paragraph{Parameters.}
Set
\begin{equation} \label{eq:eps_def}
    \epsilon \coloneqq \frac{1}{200 \kappa^{2}} = \frac{1}{\poly(n)},
    \qquad t_0 \coloneqq 1 - \epsilon, \qquad t_1 \coloneqq 1 - \frac{\epsilon}{2},
\end{equation}
abbreviate
\begin{equation} \label{eq:deltastar}
    \delta^\star \coloneqq \frac{\sqrt{\epsilon}}{5\kappa} = \frac{1}{\poly(n)},
\end{equation}
and fix the number of clock qubits to be
\begin{equation} \label{eq:k_def}
    k \coloneqq \left\lceil \left(\frac{64\kappa}{\sqrt{\epsilon}\delta^\star}\right)^{2} \right\rceil = \poly(n).
\end{equation}
The precise form of these parameters is reverse-engineered from the analysis that follows and carries no significance of its own; each choice is justified at the point where it is used. Note that all of them are independent of $\sigma$.

\paragraph{Objective function $F$.}
Let 
\begin{equation} \label{eq:lambda_def}
    \lambda(t) \coloneqq \frac{\sqrt{1-t^2}}{2t}
\end{equation}
for $t \in (0,1]$, which is nonnegative and strictly decreasing.
Let
\begin{equation}
    \varphi(t) \coloneqq \min\left\{\max\left\{\frac{t-t_0}{t_1-t_0},\, 0\right\},\, 1\right\},
\end{equation}
so that $\varphi$ is continuous, piecewise linear, and $\frac{2}{\epsilon}$-Lipschitz, with $\varphi(t)=0$ for $t \le t_0$ and $\varphi(t)=1$ for $t \ge t_1$.
Let
\begin{equation} \label{eq:G_def}
    G(t) \coloneqq \lambda\left(\min\{\max\{t,t_0\},t_1\}\right) D(\varphi(t)), \qquad t \in [-1,1],
\end{equation}
where the argument of $\lambda$ is $t$ clamped to $[t_0,t_1]$.
Thus $G$ is continuous, with
\begin{equation} \label{eq:G_cases}
    G(t) = \left\{
        \begin{array}{ll}
            \lambda(t_0)D(0) = 0, & t \le t_0, \\[2pt]
            \lambda(t)D\left(\frac{t-t_0}{t_1-t_0}\right), & t_0 \le t \le t_1, \\[2pt]
            \lambda(t_1)D(1), & t \ge t_1 .
        \end{array}
    \right.
\end{equation}

\begin{lemma} \label{lem:G_bounds}
    $M_G \leq \sqrt{2 \epsilon} M_D$ and $L_G \leq \frac{4(L_D + M_D)}{\sqrt{\epsilon}}$.
\end{lemma}

\begin{proof}
    Without loss of generality assume $t,t' \in [t_0,t_1]$ below given \eqref{eq:G_cases}.
    As $\lambda$ is decreasing and $\epsilon \leq \frac12$ by definition, we have $\lambda(t) \leq \lambda(t_0) = \frac{\sqrt{1-(1-\epsilon)^2}}{2(1-\epsilon)} \leq \frac{\sqrt{2\epsilon}}{2 \cdot \frac12} = \sqrt{2 \epsilon}$.
    Similarly $\abs{\lambda'(t)} = \frac{1}{2t^2\sqrt{1-t^2}} \leq \frac{1}{2t_0^2\sqrt{1-t_1^2}} \leq \frac{1}{2(1/2)^2\sqrt{1-(1-\epsilon/2)^2}} \leq \frac{1}{\frac12 \sqrt{\frac78 \epsilon}} \leq \frac{4}{\sqrt{\epsilon}}$.

    Therefore $\norm{G(t)} \leq \abs{\lambda(t)} \cdot \norm{D(\varphi(t))} \leq \lambda(t_0) M_D \leq \sqrt{2 \epsilon} M_D$, which proves the first claim.
    For the second, using that $\varphi$ is $\frac{2}{\epsilon}$-Lipschitz,
    \begin{equation}
        \begin{aligned}
            \norm{G(t) - G(t')} &= \norm{\lambda(t) D(\varphi(t)) - \lambda(t') D(\varphi(t'))} \\
            &\le \norm{\lambda(t) D(\varphi(t)) - \lambda(t) D(\varphi(t'))} + \norm{\lambda(t) D(\varphi(t')) - \lambda(t') D(\varphi(t'))} \\
            &\le \abs{\lambda(t)} \cdot \norm{D(\varphi(t)) - D(\varphi(t'))} + \abs{\lambda(t) - \lambda(t')} \cdot \norm{D(\varphi(t'))} \\
            &\le \sqrt{2 \epsilon} \cdot \frac{2L_D}{\epsilon}\abs{t-t'} + \frac{4}{\sqrt{\epsilon}}\abs{t-t'} \cdot M_D \\
            &\leq \frac{4(L_D + M_D)}{\sqrt{\epsilon}}\abs{t-t'}.
        \end{aligned}
    \end{equation}
\end{proof}

Let $W \coloneqq \sum_{j \in [k]} I^{\otimes j-1} \otimes \ketbra{1}{1} \otimes I^{\otimes k-j}$ be the Hamming weight operator, so that $W\lrb{c} = \abs{c}$, and let $t(c) \coloneqq 1 - \frac{2\abs{c}}{k} = \frac1k \sum_{j \in [k]} \braket{c|Z_j|c} \in [-1,1]$ be the magnetization per qubit of $\ket{c}$.
Let $F$ be the diagonal Hamiltonian on $k+n$ qubits, split into a \emph{clock register} of $k$ qubits and a \emph{problem register} of $n$ qubits, defined by
\begin{equation} \label{eq:F_def}
    F\lrb{c,x} = W\lrb{c} + G(t(c))\lrb{x}, \qquad c \in \{0,1\}^k,\ x \in \{0,1\}^n .
\end{equation}

Since $k = \poly(n)$, the following is obvious by the explicit formulas for $F$ in \eqref{eq:F_def} and $G$ in \eqref{eq:G_def}.

\begin{fact} \label{lem:F_oracle}
    One query to $F$ can be simulated using $O(1)$ queries to $D$ in $\poly(n)$ time without any information about $\sigma$.
\end{fact}

\subsection{Spectral Gap Analysis} \label{sec:spectral_gap}

The adiabatic path we analyze is
\begin{equation} \label{eq:H_def}
    \begin{aligned}
        H(s) &= - (1 - s) X_{[k+n]} + s F \\
        &= \underbrace{\left(- (1 - s) X_{[k]} + s W\right)}_{\eqqcolon H_\clock(s)} \otimes I - (1 - s) I \otimes X_{[n]} + s \underbrace{\sum_{\substack{c \in \{0,1\}^k \\ x \in \{0,1\}^n}} G(t(c)) \lrb{x}\ket{c,x}\bra{c,x}}_{\eqqcolon \widehat{G}} ,
    \end{aligned}
\end{equation}
for $s \in [0,1]$, where we used $X_{[k+n]} = X_{[k]} \otimes I + I \otimes X_{[n]}$ and \eqref{eq:F_def}.
Note that $\widehat{G} = \sum_{c} \ketbra{c}{c} \otimes G(t(c))$ is block diagonal in the clock register, so
\begin{equation} \label{eq:Ghat_norm}
    \norm{\widehat{G}} =  \max_{c \in \{0,1\}^k} \norm{G(t(c))} \leq M_G.
\end{equation}

We introduce notation that serves to state \Cref{lem:clock_spec}:
For $s \in [0,1]$, set
\begin{equation} \label{eq:Delta_m_def}
    \Delta(s) \coloneqq \sqrt{s^2 + 4(1-s)^2} = \sqrt{5\left(s-\frac45\right)^2 + \frac45} \geq \frac{2}{\sqrt5}, \qquad m(s) \coloneqq \frac{s}{\Delta(s)},
\end{equation}
and define the normalized state
\begin{equation} \label{eq:psi_def}
    \ket{\psi(s)} \coloneqq \frac{\sqrt{\Delta(s)+s}\ket{0} + \sqrt{\Delta(s)-s}\ket{1}}{\sqrt{2\Delta(s)}},
    \qquad
    \ket{\psi_\clock(s)} \coloneqq \ket{\psi(s)}^{\otimes k}.
\end{equation}

\begin{lemma} \label{lem:clock_spec}
    For every $s \in [0,1]$, the Hamiltonian $H_\clock(s)$ has ground state $\ket{\psi_\clock(s)}$ and spectral gap $\delta_{H_\clock(s)} = \Delta(s)$, and the magnetization of $\ket{\psi(s)}$ is $\braket{\psi(s) | Z | \psi(s)} = m(s)$.
\end{lemma}

\begin{proof}
    Note that $H_\clock(s)$ is the Kronecker sum of $k$ copies of the single-qubit Hamiltonian
    \begin{equation}
        A \coloneqq -(1-s) X + s\ketbra{1}{1} = \begin{pmatrix} 0 & -(1-s) \\ -(1-s) & s \end{pmatrix}.
    \end{equation}
    Diagonalizing $A$ gives eigenvalues $\frac{s \pm \Delta(s)}{2}$ with ground state $\ket{\psi(s)}$, so $\delta_A = \Delta(s)$.

    By \Cref{fact:kron}, the spectrum of $H_\clock(s)$ is the $k$-fold Minkowski sum of the spectrum of $A$, with the corresponding tensor products as eigenvectors;
    hence $\delta_{H_\clock(s)} = \delta_A = \Delta(s)$ and the ground state is $\ket{\psi_\clock(s)} = \ket{\psi(s)}^{\otimes k}$.
    Finally, $\braket{\psi(s)|Z|\psi(s)} = \frac{(\Delta(s)+s)-(\Delta(s)-s)}{2\Delta(s)} = m(s)$.
\end{proof}

Since $\ket{\psi_\clock(s)} = \ket{\psi(s)}^{\otimes k}$, \Cref{lem:clock_spec} says that $m(s)$ is the magnetization per qubit of the clock ground state $\ket{\psi_\clock(s)}$. 
The following fact is immediate from the definitions of $m$ in \eqref{eq:Delta_m_def} and $\lambda$ in \eqref{eq:lambda_def}.

\begin{fact} \label{fact:m_facts}
    The function $m \colon [0,1] \to [0,1]$ is a continuous, strictly increasing bijection, and for every $s \in [0,1]$,
    \begin{equation} \label{eq:one_minus_s}
        1-s = \frac{\sqrt{1-m(s)^2}}{2m(s) + \sqrt{1-m(s)^2}} .
    \end{equation}
    Moreover $\lambda(m(s)) = \frac{1-s}{s}$, equivalently $s\lambda(m(s)) = 1-s$, for every $s \in (0,1]$.
\end{fact}


By \eqref{eq:H_def}, the only term that couples the clock ground state to the rest of the clock spectrum is $s\widehat{G}$, whose norm is deliberately made small compared to the clock gap $\delta_{H_\clock} = \inf_s \Delta(s) = \frac{2}{\sqrt5}$.
Indeed, recall from \Cref{sec:idea} that $F$ is designed around the following idealized picture:
Fixing the clock register in the ground state $\ket{\psi_\clock(s)}$ of $H_\clock(s)$ turns the adiabatic path $H(s)$ into an effective Hamiltonian on the problem register,
\begin{equation}
    \left(\bra{\psi_\clock(s)} \otimes I\right) H(s) \left(\ket{\psi_\clock(s)} \otimes I\right).
\end{equation}
Measuring the clock in the computational basis gives a string $c$ whose magnetization per qubit $t(c)$ equals $m(s)$ on average.
Were $t(c)$ exactly $m(s)$, the effective Hamiltonian would be, up to a constant shift,
\begin{equation}
    -(1-s)X_{[n]} + sG(m(s)) = \left\{
        \begin{array}{ll}
            (1-s)K(\varphi(m(s))), & m(s) \in [0,t_1], \\
            -(1-s)X_{[n]} + s\lambda(t_1)D(1), & m(s) \in (t_1,1],
        \end{array}
    \right.
\end{equation}
where $K$ is the hard path from \Cref{thm:cited} (see \eqref{eq:Kbar_hard_path} for the first case).
The second case $m(s) \in (t_1,1]$ is built into $G$ solely to keep the effective Hamiltonian from vanishing as $s \to 1$, which would close its spectral gap.

Of course, $H(s)$ only approximates the picture above, so we need to bound the resulting error quantitatively.
The next three definitions formalize the argument into a chain of idealizations of $H(s)$: we first pin the clock to $\ket{\psi_\clock(s)}$, then replace its fluctuating magnetization $t(c)$ by the mean $m(s)$.
\Cref{lem:Kbar_gap} bounds the gap of the fully idealized Hamiltonian, and \Cref{lem:Kstar_gap,lem:H_gap} then bound the error incurred by each idealization in turn.

Let $\widetilde{G}(s)$ be obtained from $\widehat{G}$ by fixing the clock register in its ground state $\ket{\psi_\clock(s)}$:
\begin{equation}
    \begin{aligned} \label{eq:Gtilde_def}
        \widetilde{G}(s) &\coloneqq \left(\bra{\psi_\clock(s)} \otimes I\right) \widehat{G} \left(\ket{\psi_\clock(s)} \otimes I  \right).
    \end{aligned}
\end{equation}
Write $c \sim \ket{\psi_\clock(s)}$ for the random string obtained by measuring $\ket{\psi_\clock(s)}$ in the computational basis.
Since $\widehat{G} = \sum_{c} \ketbra{c}{c} \otimes G(t(c))$ is block diagonal in the clock register, we have
\begin{equation} \label{eq:Gtilde_avg}
    \widetilde{G}(s) = \expt_{c \sim \ket{\psi_\clock(s)}}[G(t(c))] .
\end{equation}
Let $\widetilde{K}(s)$ be the resulting effective Hamiltonian on the problem register:
\begin{equation}
    \begin{aligned} \label{eq:Ktilde_def}
        \widetilde{K}(s) &\coloneqq -(1 - s)X_{[n]} + s\widetilde{G}(s) = -(1 - s)X_{[n]} + s\expt_{c}[G(t(c))].
    \end{aligned}
\end{equation}
Finally, let $\overline{K}(s)$ be obtained from $\widetilde{K}(s)$ by evaluating $G$ at the mean magnetization per qubit $m(s)$ instead of averaging over $c$:
\begin{equation}
    \begin{aligned}
        \overline{K}(s) &\coloneqq -(1 - s)X_{[n]} + sG(m(s)).
    \end{aligned}
\end{equation}

\begin{lemma} \label{lem:Kbar_gap}
    $\delta_{\overline{K}} \geq \delta^\star = \frac{\sqrt{\epsilon}}{5\kappa}$.
\end{lemma}

\begin{proof}
    Fix $s \in [0,1]$. We bound $\delta_{\overline{K}(s)}$ from below by a quantity independent of $s$. Assume $s > 0$ from now on as the case $s = 0$ is trivial.

    Write $m \coloneqq m(s)$.
    By \eqref{eq:one_minus_s}, $1-s = \frac{\sqrt{1-m^2}}{2m+\sqrt{1-m^2}}$, whose denominator is at most $3$ because $0 \leq m \leq 1$.
    Hence, whenever $m \leq t_1$,
    \begin{equation} \label{eq:one_minus_s_bound}
        1 - s \geq \frac{\sqrt{1 - t_1^2}}{3} \geq \frac{\sqrt{1 - t_1}}{3} = \frac{\sqrt{\epsilon/2}}{3} \geq \frac{\sqrt{\epsilon}}{5}.
    \end{equation}

    Assume first that $m \in (0, t_1]$.
    Then $G(m) = \lambda(m)D(\varphi(m))$ by \eqref{eq:G_cases}.
    Therefore,
    \begin{equation} \label{eq:Kbar_hard_path}
        \begin{aligned}
            \overline{K}(s) &= -(1 - s)X_{[n]} + s \cdot \lambda(m) D(\varphi(m)) \\
            &= -(1 - s)X_{[n]} + s \cdot \frac{1-s}{s} D(\varphi(m)) \\
            &= (1 - s) \left( -X_{[n]} + D(\varphi(m)) \right) = (1 - s) K(\varphi(m)).
        \end{aligned}
    \end{equation}
    Hence, $\delta_{\overline{K}(s)} = (1-s)\delta_{K(\varphi(m))} \geq (1-s)\delta_K \geq \frac{1-s}{\kappa}$ by \eqref{eq:kappa_def}, which is at least $\frac{\sqrt{\epsilon}}{5\kappa} = \delta^\star$ by \eqref{eq:one_minus_s_bound}.

    Now assume $m \in (t_1, 1]$, and let $s_1$ be the unique point with $m(s_1) = t_1$ in light of \Cref{fact:m_facts}. 
    By \eqref{eq:G_cases} and \Cref{thm:cited}(a), $G(m) = \lambda(t_1)D(1) = -3n\lambda(t_1)\ketbra{u}{u}$, so
    \begin{equation}
        \overline{K}(s) = -(1 - s)X_{[n]} - 3n s \lambda(t_1) \ketbra{u}{u} .
    \end{equation}
    The second term has spectral gap $3ns\lambda(t_1) \geq 3ns_1\lambda(t_1) = 3n(1-s_1)$, where we used $s > s_1$ and then \Cref{fact:m_facts} at $s_1$.
    Since $\norm{(1-s)X_{[n]}} = (1-s)n \leq (1-s_1)n$, \Cref{fact:weyl} gives
    \begin{equation}
        \delta_{\overline{K}(s)} \geq 3n(1-s_1) - 2(1-s_1)n = (1 - s_1) n \geq \frac{n \sqrt{\epsilon}}{5} \geq \frac{\sqrt{\epsilon}}{5\kappa} = \delta^\star
    \end{equation}
    by applying \eqref{eq:one_minus_s_bound} at $s_1$, where $m(s_1) = t_1$.
\end{proof}

\begin{lemma} \label{lem:Kstar_gap}
    $\delta_{\widetilde{K}} \geq \delta_{\overline{K}} - \frac{2 L_G}{\sqrt{k}}$.
\end{lemma}

\begin{proof}
    It suffices to show $\delta_{\widetilde{K}(s)} \geq \delta_{\overline{K}(s)} - \frac{2 L_G}{\sqrt{k}}$ for each fixed $s \in [0,1]$ and then take the infimum over $s$.

    The clock state $\ket{\psi_\clock(s)}$ is a product state by \eqref{eq:psi_def}, so the bits $c_1,\ldots,c_k$ of $c \sim \ket{\psi_\clock(s)}$ are i.i.d., with $\expt[1-2c_j] = \braket{\psi(s)|Z|\psi(s)} = m(s)$ by \Cref{lem:clock_spec}.
    As $t(c) = 1-\frac{2\abs{c}}{k} = \frac1k\sum_{j \in [k]}(1-2c_j)$ is the average of $k$ i.i.d.\ random variables taking values in $\{-1,1\}$, we have
    \begin{equation} \label{eq:t_var}
        \expt_{c}[t(c)] = m(s), \qquad \var_{c}[t(c)] = \frac{1-m(s)^2}{k} \leq \frac{1}{k},
    \end{equation}
    and so $\expt_{c}[\abs{t(c)-m(s)}] \leq \sqrt{\var_{c}[t(c)]} \leq \frac{1}{\sqrt{k}}$ by Cauchy--Schwarz.
    Therefore, by \eqref{eq:Gtilde_avg},
    \begin{equation}
        \begin{aligned}
            \norm{\widetilde{G}(s) - G(m(s))} &= \norm{\expt_{c}[G(t(c))] - G(m(s))} \\
            &\leq \expt_{c}[\norm{G(t(c)) - G(m(s))}] \\
            &\leq \expt_{c}[L_G\abs{t(c) - m(s)}] \\
            &\leq L_G \expt_{c}[\abs{t(c) - m(s)}] \\
            &\leq \frac{L_G}{\sqrt{k}}.
        \end{aligned}
    \end{equation}
    Since $\widetilde{K}(s) - \overline{K}(s) = s\left(\widetilde{G}(s) - G(m(s))\right)$ and $s \leq 1$, this gives $\norm{\widetilde{K}(s) - \overline{K}(s)} \leq \frac{L_G}{\sqrt{k}}$, and the claim follows by \Cref{fact:weyl}.
\end{proof}

It remains to transfer the gap of the effective Hamiltonian $\widetilde{K}$ back to the full Hamiltonian $H$.
We address this via a block decomposition induced by the clock ground state $\ket{\psi_\clock(s)}$, together with the following elementary bound, which plays the same role as the first-order perturbation theory argument in~\cite{leng2025subexponentialquantumspeedupoptimization}.

\begin{lemma} \label{lem:block_gap}
    Let $\mathcal{H} = \mathcal{P} \oplus \mathcal{Q}$ be a finite-dimensional Hilbert space with $\dim\mathcal{P} \geq 2$, and let $H$ be Hermitian on $\mathcal{H}$ with block form
    \begin{equation}
        H = \begin{pmatrix} A & R^{\dagger} \\ R & B \end{pmatrix}
    \end{equation}
    with respect to this decomposition.
    If $\gamma \coloneqq \mu_0(B) - \mu_0(A) > 0$, then
    \begin{equation}
        \delta_H \geq \min\left\{\delta_A, \gamma\right\} - 2\norm{R} .
    \end{equation}
\end{lemma}

\begin{proof}
    Let $H_{\mathrm{bd}} \coloneqq \begin{pmatrix} A & 0 \\ 0 & B \end{pmatrix}$ be the block-diagonal part of $H$, whose spectrum is the union of the spectra of $A$ and $B$, counted with multiplicity.
    Since $\mu_0(A) < \mu_0(B)$ and $\dim\mathcal{P} \geq 2$, we have $\mu_0(H_{\mathrm{bd}}) = \mu_0(A)$ and $\mu_1(H_{\mathrm{bd}}) = \min\{\mu_1(A), \mu_0(B)\}$, so
    \begin{equation}
        \delta_{H_{\mathrm{bd}}} = \min\left\{\mu_1(A) - \mu_0(A), \mu_0(B) - \mu_0(A)\right\} = \min\left\{\delta_A, \gamma\right\} .
    \end{equation}
    The remainder $H - H_{\mathrm{bd}} = \begin{pmatrix} 0 & R^{\dagger} \\ R & 0 \end{pmatrix}$ has norm $\norm{R}$, so $\norm{H - H_{\mathrm{bd}}} = \norm{R}$.
    The claim then follows by \Cref{fact:weyl}.
\end{proof}

\begin{lemma} \label{lem:H_gap}
    Assume $M_G \leq \frac{1}{10}$. Then $\delta_H \geq \min\left\{\delta_{\widetilde{K}}, \frac23\right\} - \frac{2L_G}{\sqrt{k}}$.
\end{lemma}

\begin{proof}
    Fix $s \in [0,1]$, and split the Hilbert space according to the clock ground state: 
    \begin{equation}
        P \coloneqq \ketbra{\psi_\clock(s)}{\psi_\clock(s)} \otimes I, \qquad Q \coloneqq I - P.
    \end{equation}
    Let $\mathcal{P}$ and $\mathcal{Q}$ be the spaces onto which $P$ and $Q$ project, respectively.
    By \eqref{eq:H_def}, write $H(s) = H_0(s) + s\widehat{G}$ with
    \begin{equation}
        H_0(s) \coloneqq H_\clock(s) \otimes I - (1-s) I \otimes X_{[n]},
    \end{equation}
    and write
    \begin{equation}
        H(s) = \begin{pmatrix} A & R^{\dagger} \\ R & B \end{pmatrix}
    \end{equation}
    as in \Cref{lem:block_gap}, where $A \coloneqq P H(s) P$, $B \coloneqq Q H(s) Q$, and $R \coloneqq Q H(s) P$.

    \medskip\noindent
    \emph{The block $R$ exactly captures the fluctuation of $t(c)$.}
    Note that $\mathcal{P}$ is invariant under $H_\clock(s) \otimes I$ and $I \otimes S$ for any matrix $S$.
    Hence $H_0(s)$ is block diagonal, so $R = sQ\widehat{G}P$. Moreover, $Q\left(I \otimes G(m(s))\right)P = 0$, so
    \begin{equation}
        R = sQ\left(\widehat{G} - I \otimes G(m(s))\right)P .
    \end{equation}
    Thus, estimating the norm of $R$ reduces to estimating that of $\left(\widehat{G} - I \otimes G(m(s))\right)\left(\ket{\psi_\clock(s)} \otimes \ket{v}\right)$ for arbitrary test state $\ket{\psi_\clock(s)} \otimes \ket{v}$ with $\ket{v} \in \mathbb{C}^{2^n}$.
    Write $\ket{\psi_\clock(s)} = \sum_{c} \alpha_c \ket{c}$. 
    Note that $c \sim \ket{\psi_\clock(s)}$ takes each value $c$ with probability $\abs{\alpha_c}^2$.
    Since $\widehat{G} = \sum_{c} \ketbra{c}{c} \otimes G(t(c))$, we have
    \begin{equation}
        \begin{aligned}
            \norm{\left(\widehat{G} - I \otimes G(m(s))\right)\left(\ket{\psi_\clock(s)} \otimes \ket{v}\right)}^2 
            &= \norm{\sum_{c} \alpha_c \ket{c} \otimes \left(G(t(c)) - G(m(s))\right)\ket{v}}^2 \\
            &= \sum_{c} \abs{\alpha_c}^2 \norm{\left(G(t(c)) - G(m(s))\right)\ket{v}}^2 \\
            &= \expt_{c}\left[\norm{\left(G(t(c)) - G(m(s))\right)\ket{v}}^2\right] \\
            &\leq L_G^2 \expt_{c}\left[\left(t(c) - m(s)\right)^2\right] \norm{\ket{v}}^2 \\
            &\leq \frac{L_G^2}{k} \norm{\ket{v}}^2,
        \end{aligned}
    \end{equation}
    where the last step uses $\expt_{c}\left[\left(t(c) - m(s)\right)^2\right] = \var_{c}[t(c)] \leq \frac1k$ by \eqref{eq:t_var}.
    Since $s \leq 1$ and $\norm{Q} \leq 1$, we conclude
    \begin{equation} \label{eq:R_norm}
        \norm{R} \leq \frac{L_G}{\sqrt{k}} .
    \end{equation}

    \medskip\noindent
    \emph{The block $A$ is $\widetilde{K}(s)$ up to a shift.}
    Identify $\mathcal{P}$ with $\mathbb{C}^{2^n}$ via the isometry $\ket{x} \mapsto \ket{\psi_\clock(s)} \otimes \ket{x}$, and let $E_0$ be the ground energy of $H_\clock(s)$.
    Recall from \eqref{eq:Gtilde_def} and \eqref{eq:Ktilde_def} that
    \begin{equation}
        \widetilde{G}(s) = \left(\bra{\psi_\clock(s)} \otimes I\right) \widehat{G} \left(\ket{\psi_\clock(s)} \otimes I\right), \qquad \widetilde{K}(s) = -(1 - s)X_{[n]} + s\widetilde{G}(s) .
    \end{equation}
    Applying the map $S \mapsto PSP$ to each of the three terms of $H(s)$ in \eqref{eq:H_def}, then identifying $\mathcal{P}$ with $\mathbb{C}^{2^n}$ as above, gives $E_0 I$, $-(1-s)X_{[n]}$, and $s\widetilde{G}(s)$ respectively, so
    \begin{equation} \label{eq:A_shift}
        A = E_0 I + \widetilde{K}(s),
        \qquad\text{hence}\qquad
        \delta_A = \delta_{\widetilde{K}(s)} .
    \end{equation}

    \medskip\noindent
    \emph{The ground energies of $A$ and $B$ are separated.}
    Let $E_0 \leq E_1 \leq \cdots$ be the eigenvalues of $H_\clock(s)$.
    By the definition $H_0(s) = H_\clock(s) \otimes I - (1-s) I \otimes X_{[n]}$, it is straightforward that
    \begin{equation}
        \mu_0\left(H_0(s)|_{\mathcal{P}}\right) = E_0 - (1-s)n,
        \qquad
        \mu_0\left(H_0(s)|_{\mathcal{Q}}\right) = E_1 - (1-s)n = E_0 + \delta_{H_\clock(s)} - (1-s)n .
    \end{equation}
    Since $A$ and $B$ are obtained from $H_0(s)|_{\mathcal{P}}$ and $H_0(s)|_{\mathcal{Q}}$ by adding $sP\widehat{G}P|_{\mathcal{P}}$ and $sQ\widehat{G}Q|_{\mathcal{Q}}$, each of norm at most $sM_G \leq M_G$, \Cref{fact:weyl} gives
    \begin{equation} \label{eq:gamma_bound}
        \gamma \coloneqq \mu_0(B) - \mu_0(A) \geq \delta_{H_\clock(s)} - 2M_G \geq \frac{2}{\sqrt 5} - \frac15 \geq \frac23,
    \end{equation}
    where $\frac{2}{\sqrt 5}$ comes from \Cref{lem:clock_spec} and \eqref{eq:Delta_m_def}, and $\frac15$ comes from the assumption.

    \medskip\noindent
    \emph{Putting everything together.}
    Applying \Cref{lem:block_gap}, we have
    \begin{equation}
        \delta_{H(s)} \geq \min\left\{\delta_{A},\gamma\right\} - 2\norm{R} \overset{\eqref{eq:A_shift}}{=} \min\left\{\delta_{\widetilde{K}(s)},\gamma\right\} - 2\norm{R}
        \overset{\eqref{eq:gamma_bound}\eqref{eq:R_norm}}{\geq} \min\left\{\delta_{\widetilde{K}(s)},\frac23\right\} - \frac{2L_G}{\sqrt{k}}.
    \end{equation}
\end{proof}

\begin{lemma} \label{lem:H_total_gap}
    $M_G \leq \frac{1}{10}$, and hence $\delta_H \geq \frac{\delta^\star}{2} = \frac{\sqrt{\epsilon}}{10\kappa} \geq \frac{1}{\poly(n)}$.
\end{lemma}

\begin{proof}
    Applying the bound of $M_G$, the definition of $\epsilon$, and the definition of $\kappa$ sequentially from \Cref{lem:G_bounds}, \eqref{eq:eps_def}, and \eqref{eq:kappa_def}, we have
    \begin{equation}
        M_G \leq \sqrt{2\epsilon}M_D \leq \frac{M_D}{10\kappa} \leq \frac{1}{10} .
    \end{equation}
    Moreover, the definition of $k$, the bound of $L_G$, and the definition of $\kappa$ from \eqref{eq:k_def}, \Cref{lem:G_bounds}, and \eqref{eq:kappa_def} give $\sqrt{k} \geq \frac{8}{\delta^\star} \cdot \frac{8\kappa}{\sqrt{\epsilon}} \geq \frac{8L_G}{\delta^\star}$, which implies $\frac{2L_G}{\sqrt{k}} \leq \frac{\delta^\star}{4}$.
    Applying \Cref{lem:Kstar_gap,lem:Kbar_gap} in order we then have
    \begin{equation}
        \delta_{\widetilde{K}} \geq \delta_{\overline{K}} - \frac{2L_G}{\sqrt{k}} \geq \delta^\star - \frac{\delta^\star}{4} = \frac{3\delta^\star}{4} .
    \end{equation}
    Now, $\delta^\star = \frac{\sqrt{\epsilon}}{5\kappa} \leq \frac{1}{5\sqrt{2}} < \frac89$ by $\epsilon \leq \frac12$ and $\kappa \geq 1$, so $\frac{3\delta^\star}{4} < \frac23$ and therefore $\min\left\{\delta_{\widetilde{K}},\frac23\right\} \geq \frac{3\delta^\star}{4}$.
    Since $M_G \leq \frac{1}{10}$, \Cref{lem:H_gap} applies:
    \begin{equation}
        \delta_H \geq \min\left\{\delta_{\widetilde{K}},\frac23\right\} - \frac{2L_G}{\sqrt{k}} \geq \frac{3\delta^\star}{4} - \frac{\delta^\star}{4} = \frac{\delta^\star}{2} .
    \end{equation}
    As $\kappa = \poly(n)$ by \eqref{eq:kappa_def} and $\epsilon = \frac{1}{\poly(n)}$ by \eqref{eq:eps_def}, we have $\frac{\delta^\star}{2} = \frac{\sqrt{\epsilon}}{10\kappa} = \frac{1}{\poly(n)}$ as claimed.
\end{proof}

\subsection{Proof of the Main Theorem} \label{sec:main_theorem}

\begin{theorem} \label{thm:main}
    For every integer $N \geq 1$, there exists a family $\{F_\sigma\}_\sigma$ of diagonal $N$-qubit Hamiltonians with $\norm{F_\sigma} \leq \poly(N)$, such that the adiabatic path
    \begin{equation}
        H_\sigma(s) = -(1-s)\sum_{i \in [N]} X_i + s F_\sigma, \qquad s \in [0,1],
    \end{equation}
    has the following properties, given oracle access to $F_\sigma$.
    \begin{enumerate}[label=(\alph*)]
        \item $F_\sigma\lrb{z}$ attains its minimum at $z = z_\sigma$, which is unique by a margin of $\Omega\left(\frac{1}{\poly(N)}\right)$.
        \item $\delta_{H_\sigma} \geq \frac{1}{\poly(N)}$ and $M_{H_\sigma} \leq \poly(N)$.
        \item There is a distribution over $\sigma$ such that every classical algorithm making at most $\exp(N^{\Omega(1)})$ queries to $F_\sigma$ outputs the minimizer of $F_\sigma$ with probability at most $\exp(-N^{\Omega(1)})$.
    \end{enumerate}
\end{theorem}

\begin{remark*}
    To obtain \Cref{thm:intro}, set $n\coloneqq N$ and $f_\sigma(z) \coloneqq F_\sigma\lrb{z}$.
    We use the linear schedule $s(\tau)=\tau/T$ for $\tau\in[0,T]$ and the initial ground state $\ket{+}^{\otimes N}$.
    Since $\norm{\dot{H}_\sigma(s)}\leq N+\norm{F_\sigma}\leq\poly(N)$ and $\ddot{H}_\sigma(s)=0$, the gap bound gives polynomial adiabatic runtime~\cite[Theorem 3]{jansen2007bounds}.
    The path is $(N+1)$-sparse with known sparsity pattern, and each matrix-entry query uses $O(1)$ queries to $f_\sigma$.
    Therefore, using a known time-dependent Hamiltonian simulation algorithm~\cite[Theorem 10]{berry2020time} takes polynomial time and queries.
    For the classical hardness, taking the approximation error below the inverse-polynomial margin in (a) forces the exact minimizer, so the classical lower bound also applies to approximate minimization.
\end{remark*}

\begin{proof}
    Without loss of generality we assume $N \geq O(1)$.
    Take an admissible size $n \geq 2$ from \Cref{thm:cited}, set $N \coloneqq k+n = \poly(n)$ by \eqref{eq:k_def}, and let $F_\sigma$ be the diagonal Hamiltonian \eqref{eq:F_def} built from $K_\sigma$.

    \medskip\noindent
    \emph{Item (a).}
    \Cref{lem:H_total_gap} gives $M_G \leq \frac{1}{10} < 1$.
    If $c = 0^k$ then $\abs{c} = 0$ and $t(c) = 1 \geq t_1$, so $G(t(c)) = \lambda(t_1)D_\sigma(1) = -3n\lambda(t_1)\ketbra{u_\sigma}{u_\sigma}$ by \eqref{eq:G_cases} and \Cref{thm:cited}(a), and hence
    \begin{equation}
        F_\sigma\lrb{0^k,x} =
        \begin{cases}
            -3n\lambda(t_1) < 0, & x = u_\sigma, \\
            0, & x \neq u_\sigma .
        \end{cases}
    \end{equation}
    If instead $\abs{c} \geq 1$, then $F_\sigma\lrb{c,x} \geq \abs{c} - M_G \geq 1 - M_G > 0$.
    Thus $z_\sigma = (0^k,u_\sigma)$ is the unique minimizer of $F_\sigma$.
    Moreover, the margin at this minimum is $3 n \lambda(t_1)$, or more conveniently,
    \begin{equation}
        \delta_{F_\sigma} = \delta_{H_\sigma(1)} \geq \delta_{H_\sigma} \geq \frac{1}{\poly(N)}
    \end{equation}
    by \Cref{lem:H_total_gap}.

    \medskip\noindent
    \emph{Item (b).}
    By \Cref{lem:H_total_gap}, we have $\delta_{H_\sigma} \geq \frac{1}{\poly(N)}$.
    By the definition of $F_\sigma$ in\eqref{eq:F_def} and $M_G \leq \frac{1}{10}$, we also have
    \begin{equation}
        \norm{F_\sigma} \leq k + M_G \leq k + \frac{1}{10} \leq \poly(N), \qquad
        M_{H_\sigma} \leq \norm{X_{[N]}} + \norm{F_\sigma} = N + \norm{F_\sigma} \leq \poly(N).
    \end{equation}

    \medskip\noindent
    \emph{Item (c).}
    Choose the distribution over $\sigma$ from \Cref{thm:cited}(b), and let $\mathcal{A}$ be a classical algorithm making $q$ queries to $F_\sigma$ that outputs its minimizer with probability $p$.
    This minimizer is $(0^k,u_\sigma)$ by item (a), so $\mathcal{A}$ also finds $u_\sigma$ with probability at least $p$; and by \Cref{lem:F_oracle} each of its queries can be answered with one query to $D_\sigma$.
    This converts $\mathcal{A}$ into a classical algorithm making $q$ queries to $D_\sigma$ that finds $u_\sigma$ with probability at least $p$.
    Hence \Cref{thm:cited}(b) gives $p \leq \exp(-n^{\Omega(1)})$ whenever $q \leq \exp(n^{\Omega(1)})$.
    Since $N = \poly(n)$, both bounds have the asserted forms in $N$.

    We have proved the theorem for infinitely many $N = \poly(n)$ generated by admissible $n \geq 2$.
    For general sufficiently large $N$, choose the largest admissible $n$ with $k(n)+n \leq N$. If $n'$ is the next admissible size, then $N < k(n')+n' \leq \poly(n') \leq \poly(n)$, so the classical bounds retain their asserted forms in $N$. 
    Then, simply pad $F_\sigma$ by the Hamming weight of $N-k-n$ independent qubits.
\end{proof}

  \section*{AI Methodology}

\begin{flushleft}
\renewcommand{\arraystretch}{1.3}
\begin{tabular}{@{}l@{\qquad}c@{}}
\textbf{Activity} & \textbf{AI used} \\
Asking the research question & \texttt{No} \\
Coming up with the approach & \texttt{Yes} \\
Proof development & \texttt{Yes} \\
Writing and exposition & \texttt{No} \\
Checking for bugs & \texttt{Yes} \\
Copy editing & \texttt{Yes} \\
\end{tabular}
\end{flushleft}

\begin{sloppypar}
\paragraph{Notes.}
GPT-6 Luna suggested the main proof strategy when prompted to improve the linearization technique in~\cite{leng2025subexponentialquantumspeedupoptimization}.
GPT-6 Astra, GPT-6.1 Sol, GPT-6 Luna, and Claude Opus 5.5 were used in the preparation of the manuscript.
The author independently checked the AI-assisted arguments, reviewed the language edits, and takes sole responsibility for the correctness and presentation of the final manuscript.
\end{sloppypar}

  \printbibliography[title={References}]

\end{document}